\documentclass[12pt]{article}
\usepackage{graphics}
\usepackage{epsf}
\usepackage{amsfonts}
\usepackage{amssymb}

\catcode `@=11 \@addtoreset{equation}{section} \catcode `@=12

\def\tsector#1#2{\ {\scriptstyle #1}\hskip 1mm \mathop{\opensquare}\limits_{\lower 1mm\hbox{$\scriptstyle#2$}}^\sim\hskip 1mm}

\def\appendix{{\newpage\section*{Appendix}}\let\appendix\section%
        {\setcounter{section}{0}
        \gdef\thesection{\Alph{section}}}\section}
\def\){\right)}
\def\({\left( }
\def\]{\right] }
\def\[{\left[ }

\def\half{{1\over 2}}

\newcommand{\be}{\begin{equation}}
\newcommand{\ee}{\end{equation}}
\newcommand{\ba}{\begin{eqnarray}}
\newcommand{\ea}{\end{eqnarray}}

\def\C{{\mathbb{C}}}
\def\Z{{\mathbb{Z}}}
\begin{document}
\title{{\small \hfill SLAC-PUB-10089}~~~~~~~\\
Localized tachyon mass and a g-theorem analogue
}
\author{Sang-Jin Sin \\   \small \sl
Stanford Linear Accelerator Center, Stanford University, Stanford
CA 94305 \footnote{Work supported partially by the department of Energy under contract number DE-AC03-76SF005515.}
\\ \small \sl and \\
\small  \sl
Department of Physics, Hanyang University, Seoul, 133-791, Korea \footnote{Permanent address} 
} 
\maketitle 
\begin{abstract} 
We study the localized tachyon condensation (LTC) of non-supersymmetric orbifold backgrounds 
in their mirror Landau-Ginzburg picture. Using he existence of four copies of (2,2) worldsheet supersymmetry,
we show  at the CFT level, that  the minimal tachyon mass in twisted sectors  
 shows somewhat analogous properties of c- or g-function.
Namely,  $m := |\alpha' M^2_{min}| $  satisfies $m(M) \geq  m(D_1\oplus D_2)={\rm max} \{m(D_1),m(D_2)\}$.
  $c$- $g$- $m$- functions share the common property 
$ f(M)\geq f(D_1\oplus D_2)$ for $f=c,g,m$, although they have different behavior in detail.
\end{abstract}

\newpage
\section{Introduction}

The study of open string tachyon condensation\cite{sen} has led to many interesting consequences 
including classification of the D-brane charge by K-theory. 
While the closed string tachyon condensation involve the change of the background spacetime 
and much more difficult, if we consider the case where tachyons can be localized 
at the singularity,  one may expect the maximal analogy with the open string case.
Along this direction, the study of localized tachyon  condensation (LTC) was  considered first  in \cite{aps}
using the brane probe and renormalization group flow and by many others\cite{vafa,hkmm,dv,dab,sin,many}. The basic picture is that 
tachyon condensation induces cascade of decays of the orbifolds to  less singular ones until the spacetime 
supersymmetry is restored. Therefore the localized tachyon condensation has geometric description as the resolution 
of the  spacetime singularities. 

Soon after, Vafa\cite{vafa} considered the problem in the Landau-Ginzburg (LG) formulation using the 
Mirror symmetry and confirmed the result of \cite{aps}. 
In \cite{hkmm}, Harvey, Kutasov, Martinec and Moore studied the same problem using the RG flow as deformation 
of chiral ring  and in term of toric geometry. 
In both papers, the worldsheet N=2 supersymmetry was utilized in essential ways. 

The tachyon condensation process can be regarded as a RG-flow, 
along which there is a decreasing quantity, c-function,
 \cite{zam} for unitary conformal field theories.
For the localized tachyon condensation in non-compact space, however,
$c$ is constant\cite{aps,hkmm} and can not measure any dynamical process.
Therefore it would be very interesting to have a quantity which has a property of monotonicity along the RG-flow. 
Along this line, the authors of \cite{hkmm} suggested a quantity, 
$g_{cl}$, which is a closed string analogue of the ground state degeneracy  in open string theory\cite{affleck}. 
On the other hand, Dabholkar and Vafa\cite{dv} suggested  the maximal R-charge of Ramond sector 
(see \cite{CV,intrilligator} for ealier study on this quantity),  
as the height of tachyon potential in the twisted sector describing the localized tachyon condensation. 
Although both suggestions are well motivated by physical intuitions, the prediction of two quantity are  
slightly different\cite{sin}. 
In \cite{sin}, it was also suggested that the lowest twisted  tachyon mass increases along RG
flow. Using the spectral flow of N=2 CFT and CTP invariance of the Ramond sector together with 
the charge-mass relation for  chiral primaries, one can easily see that 
the proposal of \cite{sin} is equivalent to the GSO projected version of the one given by \cite{dv}.

The monotonically increasing property of R-charge is related to a
theorem in singularity theory called  semi-continuity of
spectrum\cite{arnoldbk} in singularity theory, which was conjectured by
Arnold \cite{arnold} and proved later by Varchenko\cite{var} and
Steenbrink\cite{sb}.  These mathematical result  can be applied\cite{CV}
to the $N=2$ supersymmetric Landau-Ginzburg theory due to non-renormalization theorem.  
In our case,
the LG theory that is mirror to the orbifold $\C^r/\Z_n$ is not an ordinary LG theory  
but an orbifolded LG \cite{orLG} model whose chiral ring structure is very different from the ordinary LG 
and hence the theorem can not be applied directly. 
Although it is easy to see the monotonicity 
of R-charge for $\C/Z_n$ case, it is non-trivial for $\C^2/Z_n$ or higher dimensional
cases.

The main goal of this paper is to prove  that {\it the lowest tachyon mass or equivalently the minimal R-charge in orbifold CFT 
and type 0 theories increases when we compare those in UV and IR fixed points}, which is conjectured in \cite{dv,sin}. 
As shown in \cite{vafa,hkmm}, the result of decay of $\C^2/Z_n$ is a sum of two disconnected theory and this fact 
introduces extra care when we compare with initial and final theories. 
When we denote the process of a mother theory decaying into 
two disconnected daughter theories  by $M \to D_1\oplus D_2$, 
the c-function satisfies $c(M) \geq c(D_1)$ and $c(M) \geq c(D_2)$, while g-function seems to have 
$g(M)\geq g(D_1)+g(D_2)$ \cite{hkmm}. 
It turns out that for minimal tachyon mass, if we define   $m= { |\alpha' M^2_{min}| }$, it satisfies
$m(M) \geq  m(D_1\oplus D_2)={\rm max} \{m(D_1),m(D_2)\}$, 
which is again different from both c- and g-function.
Although $c$- $g$- $m$- functions have different behavior in detail they share the common property 
\be f(M)\geq f(D_1\oplus D_2), \quad f=c,g,m . \ee
In order to avoid possible confusion and also to emphasize the difference with c-theorem, we call this monotonicity 
property  of minimal tachyon mass as {\it m-theorem}.

Interestingly, our method applies only to the orbifolded Landau-Ginzburg thoery and does not apply to the generic LG theories.
In this sense, our method is complementary to the method used in mathematical literature.

We use the mirror LG picture of Vafa and the existence of the many copies of (2,2) worldsheet supersymmetries
for $\C^2/\Z_n$. In CFT of $\C^2/Z_n$, there are $2^2$ extended chiral ring structures according to the choice of 
complex structures of each $\C$ factor. We call them as $cc$, $ca$, $ac$ and $aa$ rings.  
For string spectrum, we need to put spectrum of all 4 sectors together. 
When we consider the behavior of $cc$ ring elements under the condensation of a tachyon in $cc$ ring,
we can establish an explicit mapping between spectrum  of initial and final orbifold conformal field theories.
We can show that individual R-charge of tachyons increases under the process. 
This is possible since we have control over the RG-process due to the  world sheet (2,2) 
supersymmetry off the criticality, which povide the non-renormalization theorem.
However, what happen to the R-charges of
operators in $ca$ or other  rings when a tachyon operator
in $cc$ ring condensate?  The answer is that we lose control,
since we lose all supersymmetry off the criticality hence we do not have non-renormalization theorem. \\
\noindent
What saves us from 
this difficulty is the presence of the  enhanced $2^r$ copies of (2,2) worldsheet SUSY in orbifold CFT's. 
This is because its presence allows us to choose the supersymmetry generators $G^\pm_{-\half}$ and complex structure such that 
the condensing tachyon belongs to $cc$-ring. 
We can then determine the generators of the daughter theories. 
Since we know that the final products of the decay are again orbifold theories  \cite{aps,vafa,hkmm}, 
knowing the fate of the $cc$-ring element is enough to establish the fate of entire spectrum. 
Using this special property of orbifold theories, 
we will be able to establish linear mappings for each of 4 chiral rings.  \footnote{Some of  these mapping
does not describe  tachyon condensation process of individual R-charges. They just connect between the spectrum 
of mother and daughter theories. }
We can also show that the linear mapping has the property such that the R-charges of their images are bigger than 
the R-charges  of the originals. 
The mere existence of such mappings will enable us to show our main goal: the minimal charge increases under TC. 

In section 2, we give a brief summary of mirror LG model. In section 3, we give the proof of the statement and we conclude in section 4.

\section{Mirror symmetry and Orbifolds}
In this subsection we give a summary of Vafa's work \cite{vafa} on  localized  tachyon condensation. 
The orbifold $\C^r/Z_n$ is defined by the  $\Z_n$ action given by equivalence relation 
\be
(X_1,...,X_r)\sim(\omega^{k_1}X_1,...,\omega^{k_r}X_r),\quad
\omega=e^{2\pi i /n} .\label{znaction}\ee 
We call $(k_1,\cdots, k_r)$ as the generator of the $\Z_n$ action.
The orbifold can be imbedded into the gauged linear sigma model(GLSM) \cite{wittenN2}.
The vacuum manifold of the latter  is described by the D-term constraints 
\be -n|X_0|^2+\sum_i k_i |X_i|^2=t .\ee 
Its $t\to -\infty$ limit corresponds to the orbifold and 
the $t\to\infty$ limit is the
$O(-n)$ bundle over the weighted projected space 
$WP_{k_1,...,k_r}$.
$X_0$ direction corresponds to the non-compact fiber of this bundle and $t$ plays role of 
size of the $WP_{k_1,...,k_r}$.

By dualizing  this GLSM, we get a LG model with  a superpotential\cite{HV}
\be W=\sum_{i=0}^r \exp(-Y_i), \ee
 where twisted chiral fields
$Y_i$ are periodic $Y_i\sim Y_i+2\pi i$ and  related to $X_i$ by
$Re[Y_i]=|X_i|^2.$ Introducing  the variable $ u_i:=e^{-Y_i/n},$ the D-term  constraint is expressed as 
$e^{-Y_0}=e^{t/n}\prod_i
u^{k_i} $. 
The periodicity of $Y_i$ imposes the identification : 
$u_i \sim e^{2\pi i/n} u_i $ which necessitate  modding out each $u_i$ by $\Z_n$.  
The result  is usually described by  
\be
[W=\sum_{i=1}^r u_i^{n}+e^{t/n}\prod_i u^{k_i} ]// (\Z_n)^{r-1}.\label{orLGeq}
\ee
which describe  the  mirror Landau-Ginzburg model of the linear sigma model. 
As a $t\to -\infty$ limit, mirror of the orbifold is  
\be
[W=\sum_{i=1}^r u_i^{n} ]// (\Z_n)^{r-1}. 
\ee 
 Since it is  not ordinary Landau-Ginzburg theory but an orbifolded  version,
the chiral ring structure of the theory is very different from
that of LG model. For example, the dimension of the local ring of
the super potential is always $n-1$, regardless of $r$.
  
  We list some properties of orbifolded LG theory for later use.

The true variable of the theory are $Y_i$ not $u_i$ related by $u_i=e^{-Y_i/n}$.
As a consequence,  monomial basis of the chiral ring is given by
\be
\{u_1^{p_1}u_2^{p_2}|(p_1,p_2)=(n\{jk_1/n\},n\{jk_2/n\}),
j=1,...,n-1\} \label{basis},
\ee
and $u_1^{p_1}u_2^{p_2}$ has weight $(p_1,p_2)$ and charge $(p_1/n,p_2/n)$.

\begin{itemize}
\item
It has been known (see for example \cite{sin2}, 
for $\C^2/\Z_n$, any worldsheet 
fermion generated tachyon can be constructed as a BPS state, i.e, a member of a chiral ring.
 as a consequence of existence of 4 copies of (2,2) worldsheet SUSY for this special theory.

\item
It is convenient to consider the weight of a state as sum of contribution from each complex plane. 
For example, the weight of $u_1^{na_1}u_2^{na_2}$ can be considered as sum of $a_1$ from $u_1$
and $a_2$ from $u_2$. $(a_1,a_2)$ form a point in the weight space. 
As we vary $j$ in $a_i=\{jk_i/n\}$, 
the trajectory of the point in weight space will give us a parametric plot in the plane.

\item
The weight space is a lattice in torus  of size $n\times n$. 
The identification of weights by modulo $n$ corresponds to shifting string modes.
However, periodicity of the generator $(k_1,k_2)$ is $2n$ and
$(k_1,k_2)$  and  $(k_1,k_2+n)$   do not generate the same theory in general.
We choose the standard range of $k_i$ between $-n+1$ to $n-1$.
This is because the GSO projection depends not only the R-charge
vector $(\{jk_1/n\},\{jk_2/n\})$ but also the G-parity number $G=[jk_1/n]+[jk_2/n]$.
We will comeback to  this when we discuss the GSO projection.

\item
When $n$ and $k_i$  are not relatively prime,
we have a chiral primary whose R-charge vector is $(p/n,0)$.
We call this as the reducible case and eliminate from our interests. 
This is a spectrum that is not completely localized at the tip of the orbifold.
Sometimes, even in the case we started from non-reducible theory, a tachyon condensation leads us to 
the reducible case.
\end{itemize}
 
\section{m-theorem}
The conformal 
weight in the NS-sector is related to the mass by \cite{dixon,dabholkar,lowe}
\be \frac{1}{4}\alpha'M^2= \Delta-\half.
\ee 
For $\C^1/\Z_n$ model, $\Delta_{min}=1/2n$ so that
$\alpha'M^2_{min}=-2(1-1/n)$ is proportional to the deficit angle
of the cone. The maximal R-charge in NS sector and the minimal tachyon mass is simply 
related. Let's imbed the orbifold into 8
dimensional transverse target space of lightcone  
string theory. Then the transverse spacetime is  $\C^{r}/\Z_n \times
R^{8-2r}$.
The ground states of the twisted sectors in NS sectors are
chiral or anti-chiral primary and the charge is simply given by the weight $ q=\pm 2\Delta$. 
By the spectral flow, $ q_{R}=q_{NS}-{\hat c}/2,$  
we have $ \frac{1}{4}\alpha'M^2= \half (q_{R}+\frac{{\hat c}}{2}) -\half.$
and  
\be \alpha'M^2_{min}= Q^5_{min} +{\hat c}-2, \ee 
where 2 in $Q^5_{min}=2q_{R,min}$ comes from summing left and right
R-charges. 
Using the CPT symmetry  on the Ramond sector, we have $q_R^{min}=-q_R^{max}$. 
Therefore above statement can also be written as 
\be {\rm max}\left|\alpha'M^2\right|=Q^5_{max} +2- {\hat c} :=m. \ee
%
The main goal of this paper is to show that $m$ has a property that is a reminiscent of a c-function:
\be m(UV) \geq m(IR).\ee
We call this as a m-theorem to prevent possible confusion with c-theorem or g-theorem.

\subsection{proof of the theorem}
This work is heavily dependent on the result of companion paper \cite{sin2}, where details of  
transition  of spectrum  is discussed.
Our interest is how the spectrum flows under the RG-flow.
In lack of of control of off-shell theory, it is in general difficult question to address. 
However, spectrum of UV and IR theories are readily available since both
are conformal field theory. Since any tachyon generated by worldsheet fermion  
can be thought as a chiral ring element in one choice of (2,2) SUSY, 
we can assume, without loss of generality, that the condensing tachyon  with 
weight $p=(p_1,p_2)$ is an element of $cc$-ring. 
Then consider  other element in the same $cc$-ring whose weight
is $q=(q_1,q_2)$. For definiteness, let's say $q\in \Delta_-$ in the notation of 
\cite{sin2}.
The R-charge of it is $R_q=(q_1+q_2)/n$. Now after the condensation of
  $p=(p_1,p_2)$,  $q$ is moved to
$q'=T_p^-(q)$ \cite{sin2} where, 
 \be T^-_p=\pmatrix{{p_2}/{n} & -{p_1}/{n} \cr 0 & 1}, \label{Tp-}\ee
whose  R-charge is 
\be R_{q'}=(q_2-p\times q/n)/p_2. \ee 
If $p$ represents a tachyonic state, it must be below
the diagonal. Namely, \be p_1+p_2 \leq n.\ee Therefore the
difference in the R-charge  between  before and after the process
is given by \be R_{q'}-R_{q}=(n-p_1-p_2)q_2/np_2\geq 0. \label{ineq}\ee 
The same inequality holds for $q\in \Delta_+$. 
For $p,q\in ac$ ring, we can apply the same argument by replacing 
\be 
p \to {\bar p}=(p_1,n-p_2),\quad q\to {\bar q}=(q_1,n-q_2).  \ee
Therefore we arrive at the result: \\
\noindent {\it Lemma:~~} {The R-charge of a relevant chiral
primary operator increases under condensation of tachyon in the same ring.} 

The eq. (\ref{ineq}) also shows that under the
condensation of marginal operator, there is no change in R-charge
of any operator. Due to the mass-charge relation discussed before, 
we can make the same statements for the tachyon mass. 
The above statement shows that  any of the spectrum is  a candidate of
the c-function of the twisted sector.
However, this statement does not exclude the possibility of level crossing. 
That is, the ordering of the R-charge can be changed during the process.

What happen to the R-charges of
operators in $ca$ ring when a tachyon in $cc$ ring condensate?  
The answer is that we lose control, since we lose the world sheet
(2,2) super symmetry off the criticality and we
lose non-renormalization theorem.
In fact if one naively apply the mapping $T^\pm_p$ to the $ca$-ring elements,
 we get non-integer power of $u_i$'s.
Similarly, when we condense a $ca$ ring element, we lose control over the $cc$ ring spectrum.

However, when an element in $ca$ ring turn on, we have control over other $ca$ ring elements
instead. It is holomorphic and protected by worldsheet SUSY $G^+_{ca}$.
Since we have choice of selecting complex structure
in each plane independently, we can choose any combination of complex structure
to define the holomorphic co-ordinate of $\C^2$. We can
call $u_1, {\bar u_2}$ as the holomorphic co-ordinates just as we
can call $u_1, {u_2}$ as a holomorphic coordinate. As far as other
combination does not enter in the theory, operators are protected by
the worldsheet supersymmetry.

As a warm up, we first consider the case where the minimal charge of
initial and final theories belong to $cc$-ring, so that we do not need to consider other 
chiral rings.
Let $q_0$ denote the a state of minimal R-charge, namely, \be
R[q_0] \leq R[q], \quad {\rm for~~~ all} \;\; q.\ee We want to
compare the minimal charge of the initial charge and that of the
final state. Let $q'$ be a minimal charge of a final theory. There
are two theories in the final states and one choose any of it, say
up-theory. Then $q'$ should come from a $q\in \Delta_+$  such that
$q'=T_p^+(q)$. Due to the monotonicity of R-charge, we have $
R[q']>R[q]$.  On the other hand,  $R[q]$ can not be smaller than
$R[q_0]$, by definition of $q_0$. Therefore we have inequality \be
R[q^{initial}_{min}] <R[q^{final}_{min}].\ee The same inequality
holds for the down-theory as well.

Some of the relevant operators, which are precisely those in the
triangle $\triangle  BPA$, will be pushed out to irrelevant
operator after P is condensed. One may worry about the converse
possibility that some irrelevant operators of the initial theories
flow to the relevant operator. Following lemma tells us that it
does not happen.

\noindent{\it Lemma~~:} Relevant chiral primary states of final
theory comes only from the relevant ones in the initial theory.\\
\noindent {\it Proof:}  Let $q'/p_2$ be the charge of a relevant
operator in the down-theory and $q$ be its pre-image in the
original theory, i.e, $q'={T}_p^-(q)$. Our question is whether
$q'_1+q'_2<p_2$ implies $q_1+q_2<n$ or not. This can be answered
simply by calculating the inverse of ${T}_p^-$. \be q=(T_p^-)^{-1}
(q')= \frac{n}{p_2}\left(
\begin{array}{cc} 1&p_1/n\\
 0&{p_2/n}
\end{array}\right){q'_1\choose q'_2}=  {(nq'_1+p_1q'_2)/p_2 \choose
q'_2}. \ee
Now,
\be q_1+q_2=(nq'_1+q'_2(p_1+p_2))/p_2 \leq n(q'_1+q'_2)/p_2 < n.\ee
Following is an easy consequence.:
\noindent{\it Minimal R-charge of the $cc$ ring increases under  
condensation of tachyon in $cc$-ring. 
More explicitly, 
\be {\rm min}_{l=1}^{n-1}
\left(\left\{{lk_1}/{n}\right\} +\left\{{lk_2}/{n}\right\}\right)\ee 
increases under tachyon condensation.}

So far  is story of $cc$-ring. 
We need to consider all chiral rings together. So we are interested in 
the behavior of the R-charge which is smallest in the union of $cc$ ring and $ca$-ring.
To do this, we reconsider the problem of 
the fate of $ca$-ring under the condensation of tachyon in $cc$-ring.
Although we do not have any control over the flow of the $ca$ ring spectrum,
we know what is the final theory  and its total set of the spectrum. 
We ask whether any tachyon mass of the final theory can be considered as an image of 
some mapping with the property of R-charge increasing.
To do this we want to show that 
there is a map that takes the some of chiral ring of the mother theory to $ca$ or $ac$ ring of the 
daughter theories. Notice that, in general, the $ca$ ring of $n(k_1,k_2)$ is $cc$ ring of $n(k_1,-k_2)$ and the 
daughter theory has structure $p_1(k_1,p\times k/n)\oplus p_2(-p\times k/n,k_2)$.
First, the $ca$ ring of the daughter theory $p_1(k_1,p\times k/n)$ is 
$cc$-ring of $p_1(k_1,-p\times k/n)$ which is expected to be the image of the $cc$ ring of 
$n(k_1,-k_2)$ under some mapping $F_p^+$, which is not necessarily associated with physical process.
It turns out that  $F_p^+$ can be chosen as $T^{+}_{(p_1,-p_2)}$: 
\be
F_p^+({\bar q}):=T^+_{p'}({\bar q})=(q_1,p_1-p\times q/n),\label{deff+}
\ee
where ${\bar q}=(q_1,n-q_2) \in ca$ ring and ${p'}=(p_1,-p_2)$.
One can check that 
\be
R[F^+_{p}({\bar q})]>R[{\bar q}]\;\;if\;\;  {\bar q} \in ca \; ring \label{rf+}.\ee
Similarly, the $ac$ ring of the daughter theory $p_2(-p\times k/n,k_2)$ is 
$cc$ ring of $p_2(p\times k/n, k_2)$, which can be considered as the image of the $cc$ ring of 
$ n(-k_1,k_2)$ by the map $F_p^-$ defined by
\be
F_p^-({\tilde q}) :=T^-_{-p'}(n-q_1,q_2))=(p_2+p\times q/n,q_2),
\ee
where ${\tilde q}=(n-q_1,q_2) \in ac$ ring. It can be also shown that 
 \be R[F^-_{p}({\tilde q})]>R[{\tilde q}]\;\; if\;\;  {\tilde q} \in ac \; ring.\label{rf-}\ee

Now let $q'_0$ be the  tachyon with lowest mass in the daughter theory.  
Let it belong to the up-theory. Then we can without loss of generality assume that it belongs to $ca$ ring due to the 
equivalence $ca$ and $ac$ ring in their spectrum. 
(If it belongs to $cc$ ring, we have shown already  what we want to show.) 
Then $q'=F_p^+({\bar q})$ for some ${\bar q}$, which has bigger charge than the minimal charge of initial theory.
Using the property of eq. (\ref{rf+}), 
\be
R[q']\geq R[q] \geq R[q_0], 
\ee
as desired.
Similarly, if $q'$ belongs to  down theory, we can assume that it belongs to the $ac$-ring of the down theory.
Then 
Then $q'=F_p^-({\tilde q})$ for some ${\tilde q}$, which has bigger charge than the minimal charge of initial theory $q_0$.
Using the property of eq. (\ref{rf-}), 
\be
R[q']\geq R[{\tilde q}] \geq R[q_0], 
\ee
as desired.

Therefore  we proved following:
 {In the conformal field theory of orbifolds, 
    the minimal R charge of the final theory is bigger than that of the initial theory under the condensation 
  of a tachyon generated by a world sheet fermion;}
\be \mathop{\rm min}_{l=1}^{n-1} [{\rm min}\left(
\left\{{lk_1}/{n}\right\} +\left\{{lk_2}/{n}\right\}-1 , 
\left\{{lk_1}/{n}\right\} -\left\{{lk_2}/{n}\right\} \right) ]. 
\ee 
increases when we compare its value in the initial and final theories.
Equivalently, in terms of absolute values of tachyon mass, 
 \be m(IR) \geq m(UV).\ee

A few remarks are in order:
\begin{itemize}
    \item These theorems are world
sheet fact. The same statement conjectured in \cite{sin} is the GSO projected version
for which we need to take into account the GSO projection. 
However, tachyons in NS-NS sector is not projected out by the type 0 GSO projection,
so the above conclusion is true in type 0 string theory level. For type II
we will  discuss in  detail in next subsection.  

    \item There are two independent theories in the final stage of
tachyon condensation. Each theory will have its own minimal
charges. We should take the smaller of the two, since the minimal
mass of final theory is the minimal over all final spectra.
Namely,  the minimal R-charges of two theories in final stage are not to be added
to compare with the initial one, contrary to the treatment of
$g_{cl}$ in \cite{hkmm}.
    \item  From the discussion of previous subsection, this  is the proof of the conjecture stated
by Dabholkar and Vafa in \cite{dv}. The R-charge
here is that in NS sector. The R-charge of Ramond sector is
related to that of NS sector by spectral flow. Since there are CPT invariance in Ramond sector, the
statement that minimal charge increases is equivalent to statement that maximal charge decreases.  
\end{itemize}

\section{ Discussion and Conclusion}
In this paper, we have studied the localized condensation in non-supersymmetric orbifold
using the (2,2) world sheet SUSY in  mirror LG picture. 
We study the localized tachyon condensation in Mirror Landau-Ginzburg picture as well as the
toric geometry picture of non-supersymmetric orbifold backgrounds.
Due to the two copies of (2,2) worldsheet supersymmetry,   
any worldsheet fermion generated tachyon can be considerred as a BPS state. 
Utilizing this fact, we showed that the  R-charge 
of chiral primaries increases  under the process of localized tachyon condensation. 
The minimal tachyon mass in twisted sectors increases in CFT and type 0 string and plays 
similar  role of the c- or g-function in the twisted sectors. 
By working out how the individual chiral primaries are mapped under the
tachyon condensation, we have proved that R-charges of chiral primaries increase
under tachyon condensation. We studied the GSO projection and found that
in many aspects, the separate notion of type II and type 0 is not preserved under the
tachyon condensation.

We emphasized similarity of c-,g- and m- functions in the introduction. Here we want to 
draw the reader's attention to one more object;  
the orbifold  Witten index $\mu=n$, counting the number of (un)twisted sector.
Under the transition $n(1,k)\to p_1(1,s)\oplus p_2(-s,k)$, $\mu$ obviously satisfy the 
relation 
\be
\mu(M)\geq \mu(D_1)+\mu(D_2),\ee
from the very fact that $(p_1,p_2)$ is a tachyon. This property is similar to the g-function.
Notice, however, for marginal case $\mu(M)=\mu(D_1)=\mu(D_2)$ does {\it not} hold  for witten index unlike the c-function. 
This is just parallel to the case of minimal R-charge.
Inspite of its simplicity,  we did not use this as an semi-c-object to study orbifold decay for following reasons;
\begin{itemize}
    \item It does not depends on the generator $k$. Too many theories have the same witten index. Therefore it is not a sharp indicator.
    \item It can not be defined off-shell. It is true topological index that is independent of 
    finite deformation of the theory. Therefore it can not be used for full  study of RG-flow. The minimal tachyon mass, on the other hand,
     can be defined off-shell along the line of \cite{CV,intrilligator}.
\end{itemize}   

We now discuss the limitation and related future works.
First of all, our work is confined to orbifold fixed points before and after the tachyon condensation.
It would be interesting to work out the detail of the off-shell. 
One may ask what is the geometry for the finite condensation co-efficient of LG in terms of gauged 
linear sigma model? A work related to this question has appeared \cite{kdecay}.
Another related work is \cite{minwalla}, where the Bondi 
energy \cite{bondi} as a c-function was discussed based on the earlier work by Tseytlin \cite{tseytlin2}.
The use of this work, however, seems to be rather limited for studying orbifold transitions, since it requires that the asymptotic regions
be invariant under the transition, which can not be fulfilled by the  orbifolds. 

Secondly, we considered the case where only one tachyon generated by a worldsheet fermion and 
 the method of this paper does not work if we consider simultaneous condensation of two tachyons 
in the different chiral rings. 

Thirdly, our work is mostly about CFT and type 0 theory rather than type II theory.
For type II theory, there is only one way by which the theorem can be broken, namely, 
if the  minimal charge of mother theory is projected out by the GSO   
and the minimal charge of the daughter  theory is kept, then it may happen that the minimal charge 
of the daughter theory is smaller than that of the mother theory.
It is very plausible that such possibility happens. 
Surprisingly, however, in all example we considered, the tachyon condensation that cause such possibility 
is forbidden due to GSO projection in all the examples we looked at so far. 
We wish to come back to  this issue in later publication.

\vskip 2cm
\noindent {\bf \large Acknowledgement} \\
I would like to thank  Lance Dixon, Michael Gutperle, Shamit Kachru, Amir Kashani-Poor, Matthias Klein and  Michael Peskin for discussions,
and especially to Eva Silverstein for her support and interests, and 
to Allan Adams for his collaboration in the initial stage. 
This work is partially supported by the Korea Research Foundation Grant (KRF-2002-013-D00030).

\end{document}